# Reproducing Fresnel-Arago historical experiment: a visual illustration of the main concepts of physical optics

Benoît Rogez, Thomas Chaigne, Jérôme Wenger*

*Aix Marseille Univ, CNRS, Centrale Marseille, Institut Fresnel, 13013 Marseille, France*

*\* Corresponding author: jerome.wenger@fresnel.fr*

**Abstract**

Many concepts of physical optics can be visually illustrated on a relatively simple optical setup in a table-top format, not requiring any very specific equipment. Diffraction, interferences, speckle, image formation, Fourier optics, strioscopy are among the many themes to be shown using the demonstration system described here. This short letter describes how to reproduce the setup and prepare the samples. It also gives a brief description of the experiments that can be performed to illustrate the main concepts of physical optics (except coherence). Its content should be of interest to the teachers and students at high school and early years of university.

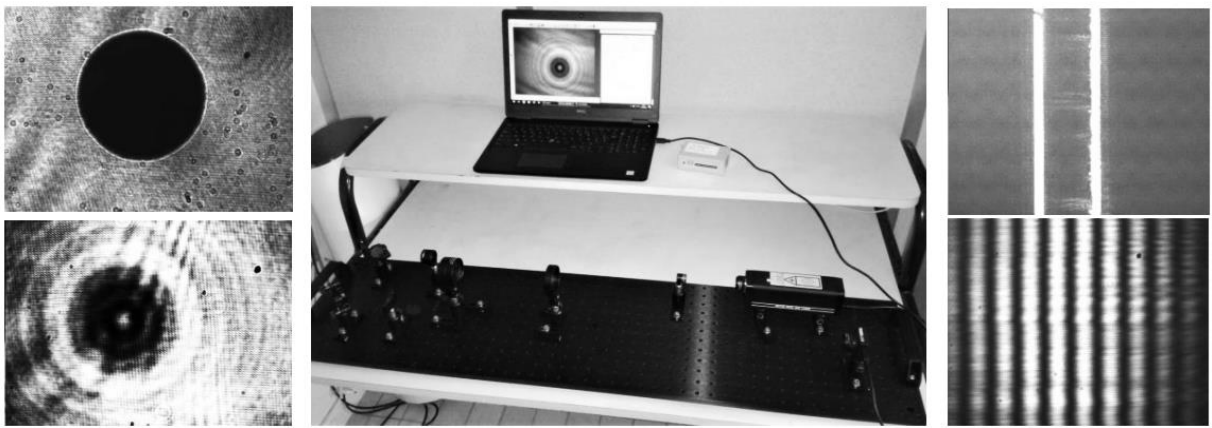

**Introduction: a brief historical background about Fresnel-Arago experiment**

At the beginning of the 19th century, the corpuscular theory of light established by Isaac Newton is striving to explain Thomas Young's observations of interference fringes when light passes through a double slits [1]. In 1818, the French Academy of Sciences launched a contest to improve the understanding of the properties of light. Augustin Fresnel sent one of the two submitted reports to the Academy of Sciences [2]. Among the reviewers was Siméon Poisson, who strongly refuted Fresnel's theory as it was considering light as a wave. Poisson then noticed that using Fresnel's theory, one should observe a bright spot in the shadow of an opaque disk [3]. According to him that was obviously ridiculous, hence Fresnel's theory was wrong. Fortunately, Augustin Fresnel was supported by François Arago, who performed the experiment suggested by Poisson. To the general surprise, he observed a bright spot in the shadow of a disk! This validated Fresnel's theory of light diffraction, which is still largely used today [4,5]. More than 200 years later, using lasers and cameras, we can easily reproduce this historical experiment which played a crucial role in our understanding of the properties of light.

**Experimental setup**

Figure 1 describes the optical setup. It is built on a Thorlabs MB30120/M breadboard measuring 1200 x 300 cm², a more compact (900 x 300 cm²) format should be achievable without much problem, or just some wood plate to cut the costs. The total weight of the apparatus is below 10 kg, and can therefore be easily carried, using for instance breadboard handles (Thorlabs BBH1). The light source is a 0.8mW HeNe laser operating at 632.8nm (Thorlabs HNLS008L-EC). The laser output is further attenuated by a neutral density filter of 0.6 optical density (Thorlabs NE06A-A) directly mounted on the laser, leading to a laser beam power of 0.2mW. This power enables a correct visualization of the light beam while being low enough to ensure correct eye safety (direct and prolonged exposure should be avoided though) without requiring any specific laser safety equipment for the teacher nor the students.

The laser beam is expanded by a telescope made of two lenses of 25 and 400 mm focal lengths (Thorlabs AC127-025-A-ML and LA1172-A-ML respectively), with a lateral expansion factor of 400/25=16x. This gives a beam diameter of about 10mm which is largely enough for the experiments. Two metal-coated mirrors then reflect the laser beam towards the sample and the camera.

The sample is hold by two clips (Thorlabs FH2 sample holder) which provides a versatile approach with enough flexibility to move the sample during the live demonstrations. The camera is a USB-powered monochrome CMOS device (Thorlabs DCC1545M) whose sensor is placed on the light beam without any further optics. A standard webcam should work (one its objective lens has been removed), yet this CMOS camera affords a simple tunability of the gain, contrast and integration time via its dedicated software supplied with it. To operate in normal light conditions, the camera is equipped with a 610 nm long pass color filter (Thorlabs FGL610M) to block most of the visible light from the room while transmitting the red laser beam.

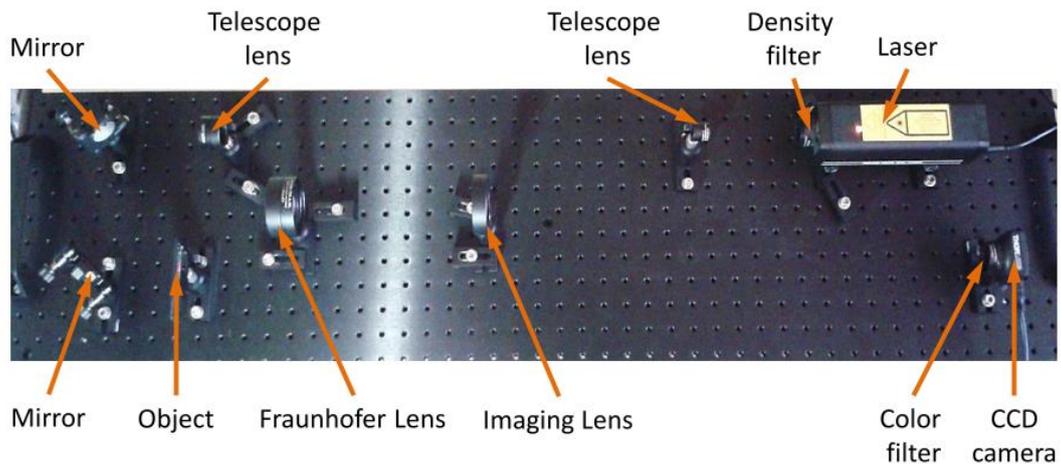

**Figure 1:** Experimental demonstration setup, see text for details

Fresnel diffraction by the sample object at finite distance is viewed directly on the camera (Fig. 2A). A supplementary lens on a flippable mount can be used to image the sample on the camera (Fig. 2B). The imaging lens is a 2 inch diameter biconvex lens of 200 mm focal length (Thorlabs LB1199-A, the 2 inch diameter just gives a better look in my opinion) mounted on a 90° flip lens mount (Thorlabs TR2F90/M). Additionally, another lens can be added to observe Fraunhofer diffraction at large distance (Fig. 2C). In this latter case, the lens is placed so that the camera stands in the image focal plane of the lens. We used a 750 mm focal length (Thorlabs AC508-750-A-ML) again on a 90° flip lens mount (Thorlabs TR2F90/M) to allow inserting and removing easily the lens without requiring optical alignment during the demonstrations to students.

The total cost of the setup if everything is bought new as described here should be around 2500€. The most expensive devices are the laser, the breadboard and the camera (about 1500€ for these three elements). The other optical elements and optomechanics are quite cheap, and the total budget can be strongly reduced if some existing laser sources, lenses or optomechanic holders can be reused.

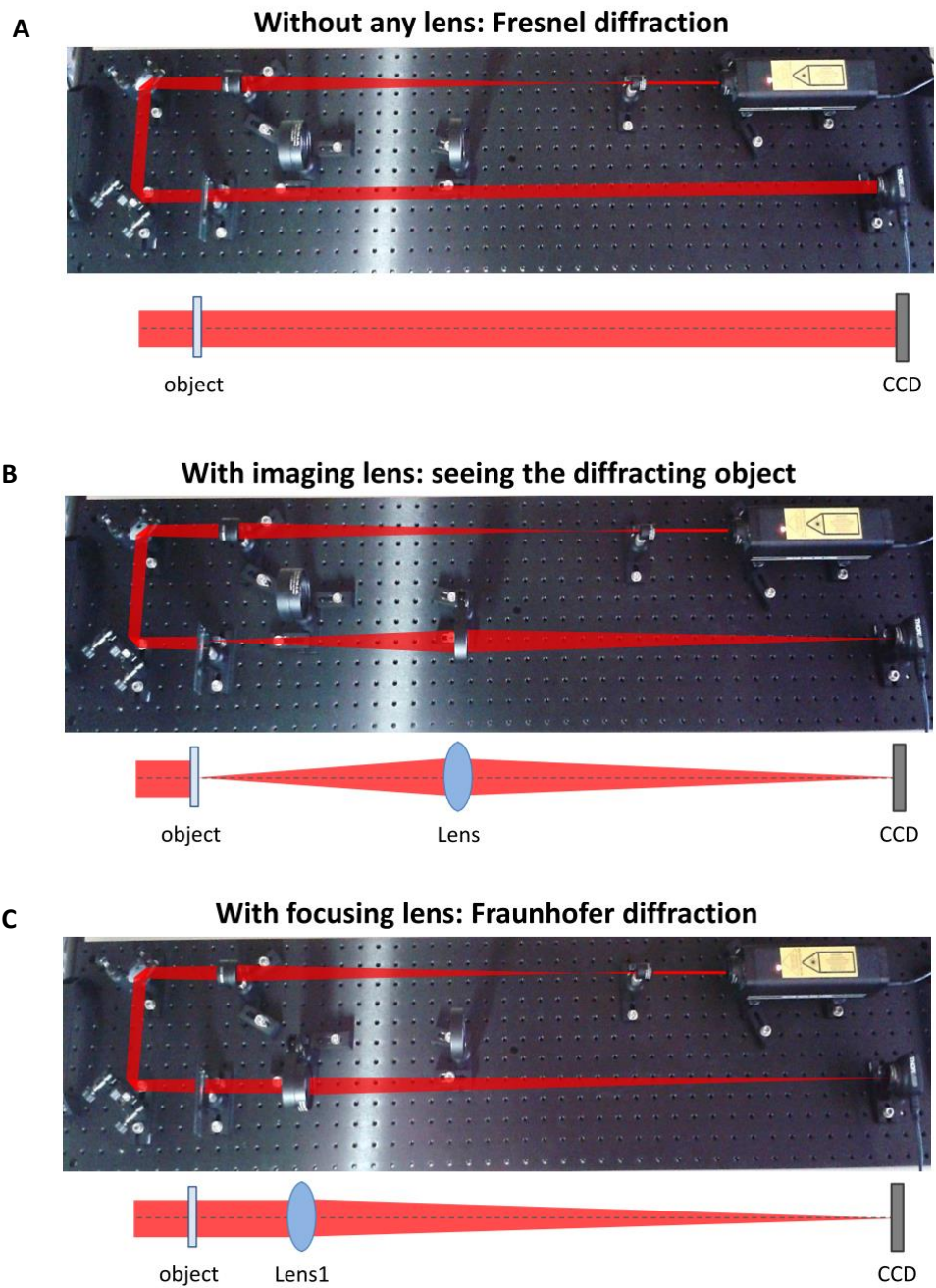

**Figure 2:** Different operation modes of the setup, for (A) direct Fresnel diffraction, (B) imaging the sample on the camera and (C) Fraunhofer diffraction. The modes A and B are the most useful for the demonstrations, the Fraunhofer diffraction (C) can be often skipped.

**Sample preparation**

One of the main challenges of this work is the sample preparation. Here we have focused on simple methods that anybody can reproduce without specific tools. Therefore, advanced micro and nanofabrication tools such as laser writing, or electron beam lithography were excluded.

To fabricate holes for the diffraction experiment, a thin plastic sheet was painted in black to make an opaque screen (using Citadel chaos black paint spray). Circular holes of diameters from 0.5 to 2 mm were then drilled manually using drill bits bought at a local hardware store.

To fabricate the opaque disks for the Fresnel-Arago spots, we deposited small dots of water-soluble black paint (Citadel Abaddon black) on a clean microscope glass slide. It takes a bit or trial and error to find the right consistency of the water–diluted black paint, but once achieved, nearly circular droplets of diameters from 0.5 to 2 mm are easily obtained. The paint can be deposited using a thin model hobbyist brush or simply the tip of a needle.

To fabricate the single and double slits for Young's interferences experiments, a cleaned microscope glass slide was covered with black paint (Citadel chaos black paint spray). Then the paint was scratched using the tip of a needle guided by a ruler. Nearly parallel slits can be obtained that way, with an interslit distance below 1 mm.

All samples are covered with a thin glass microscope coverslip to add extra mechanical protection against scratches. The sample is then sealed using cyanoacrylate glue or tape. With this protection layer, the samples could be used by several persons for many exhibitions without showing any sign of damage. However, the drawback of adding a supplementary glass coverslip is that it adds reflections and induces additional fringes on the camera image. While most audience members did not seem to notice these extra fringes, it is still worth to mention their presence.

**Experiments to be performed**

Table 1 summarizes the main experiments that can be performed on the demonstration setup. The main concepts of physical optics (diffraction, interferences, speckle, Fourier optics) are easily illustrated visually [5]. Online videos (see Supporting Media Files section below) show the experiments recorded live. The main common point between them is to frequently switch from visualizing the sample on the camera (using the imaging lens, see Fig. 2B) to observing the diffraction shadow (removing the imaging lens to be in Fresnel diffraction conditions, see Fig. 2A).

Fresnel-Arago diffraction experiment uses the black paint disks sample (Fig. 3). The main point is to observe the central bright spot in the shadow of an opaque disk. Different disk diameters can be compared. Putting the sample on a rail and moving it away from the camera gives nice results (see video online). The diffraction by a single hole is one of the most studied case (Fig. 4). Here again, the effect of different hole diameters can be illustrated. Airy diffraction pattern is well seen, and is important to discuss the resolution in optical microscopy. By moving the sample closer to the camera (~1-2cm), we can also observe dark regions at the center of the bright spot, showcasing spectacular near field diffraction patterns.

**Table 1.** Non-exhaustive list of experiments that can be performed with this demonstration system.

| Physical optics concept | Teasing text for students |
|---|---|
| Diffraction by a disk: Fresnel-Arago spot | This demonstration reproduces the historical Fresnel-Arago experiment and explores the shadow of an opaque disk, answering the question: is there light in the center of a shadow from a black disk? |
| Diffraction by a hole: the Airy disk | What is the shadow of a single hole milled in an opaque screen? The most studied case of diffraction is illustrated here experimentally using simple optical elements. |
| Young's double slits interferences | Can adding light to light create shadow? This demonstration shows that this can indeed be the case, reproducing Young's historical double slit interferences experiment, which is now the most studied interference configuration. |
| Random interferences and Speckle | What is the shadow of multiple scatterers illuminated by a coherent laser beam? This demonstration illustrates the principle of speckle formation by multiple random interferences. |
| Strioscopy and Fourier optics | Mastering optics allows you to filter an image even before it reaches the camera! This demonstration illustrates the principle of strioscopy, which is a specific application of Fourier optics. |

Young's double slit interferences is also frequently used to discuss the interference phenomenon (Fig. 5). Students can see how changing the distance between slits affects the interfringe separation.

Supplementary experiments involve speckle formation and random interferences. Here the sample consists of butterfly scales randomly deposited on a microscope glass slide. Any scatterer would do the job, using thin sand or dust should provide similar results. Similar patterns can also be easily obtained using low quality transparent pockets, the rough surface acting as a random phase mask. In this experiment the sample is placed on the rail just in front of the camera (visualizing the shadow from each individual scatterer) then translating the sample to increase the sample-camera distance, the shadows of each scatterer start to interfere until a complete random interference pattern (speckle) is obtained.

For the strioscopy experiment, the sample is a prepared microscope slide (slice of rat intestine). The image is formed on the camera, then (keeping the imaging lens) a black disk (the same as for Fresnel-Arago experiment) is inserted in the lens back focal plane (200 mm away from the lens) to block the ballistic transmitted light and filter high spatial frequencies. As a consequence, the image contrast is inverted, and fine details strike out more prominently.

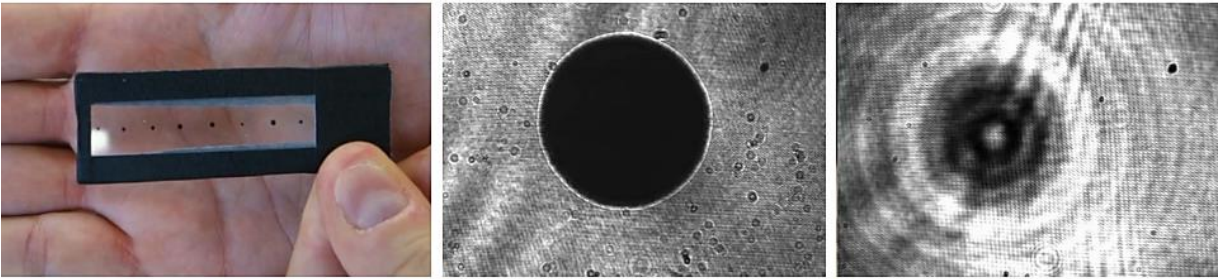

**Figure 3**. Pictures of the Fresnel-Arago diffraction experiment. (Left) black disks sample. (Center) Image of a black disk and (Right) its Fresnel diffraction pattern showing a bright spot in the shadow center.

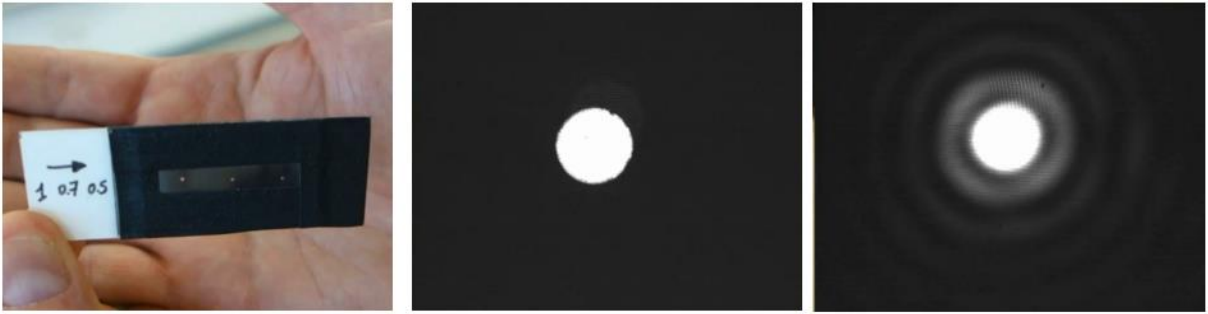

**Figure 4**. Pictures of the diffraction by a hole experiment. (Left) holes sample. (Center) Image of a single hole and (Right) its Fresnel diffraction pattern showing the Airy rings surrounding the center spot.

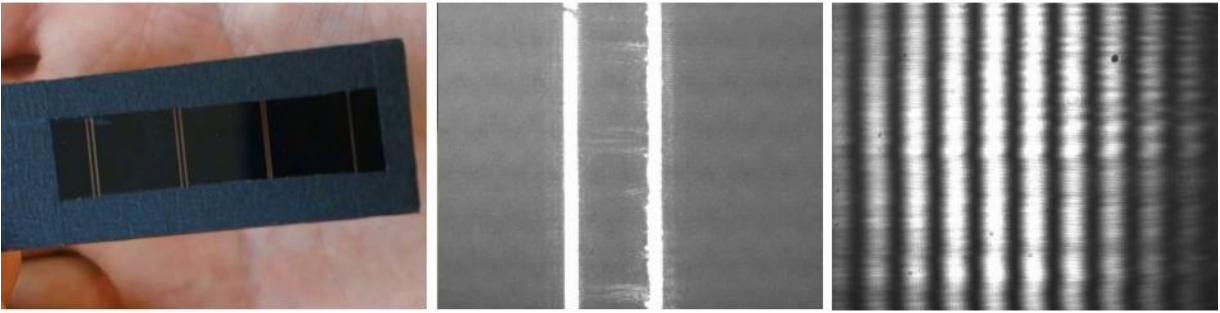

**Figure 5**. Pictures of Young's interferences experiment. (Left) Double slits sample. (Center) Image of a double slit and (Right) its interference pattern.

**Conclusions**

This note describes how to build a relatively simple optical setup to illustrate many concepts of physical optics. Diffraction, interferences, speckle, image formation, Fourier optics, strioscopy can be shown in a didactic manner. Simple tricks that can be easily reproduced are detailed for the sample preparation. In addition to the videos online (see list below), this document should help the teachers and students at high school and early years of university to reproduce such versatile demonstration system.

**Supporting media files**

The following videos illustrate the main experiments that can be performed to teach physical optics. You can also see them on the youtube channel of Institut Fresnel:

1. Arago spot
   https://youtu.be/xHHhbR5evq0

2. Optical setup description
   https://youtu.be/59M95ZwHtEk

3. Diffraction by a hole
   https://youtu.be/2SC5d9_vDPw

4. Young's double slits interferences
   https://youtu.be/8hpvyhmaLPQ

5. Speckle
   https://youtu.be/-jmv01va2pE

6. Fourier optics strioscopy
   https://youtu.be/gfouIYn_7nY

**Note.** The authors declare no competing financial interest.

**Acknowledgments.** The authors thank Brian Stout for stimulating discussions and the Institut Fresnel for providing funding for the equipment.